\documentstyle[epsf,twocolumn]{espnart}
\newcommand\prepnum{KUNS1431}
\newlength{\figwidth}
\newcommand{\Tr}{\mathop{\rm Tr}}
\makeatletter
\begin{document}
\begin{frontmatter}
\title{%
\protect\vspace*{-\baselineskip}
\vbox to3\baselineskip{\rightline{\large\sf\prepnum}}
Periodic-Orbit Bifurcation and Shell Structure
in Reflection-Asymmetric Deformed Cavity}
\author[a]{Ayumu Sugita,}
\author[b]{Ken-ichiro Arita}
\author[a]{\and Kenichi Matsuyanagi}
\address[a]{Department of Physics, Graduate School of Science,
Kyoto University, Kyoto 606-01, Japan}
\address[b]{Department of Physics, Nagoya Institute of Technology,
Nagoya 466, Japan}
\begin{abstract}
Shell structure of the single-particle spectrum for
reflection-asymmetric deformed cavity is investigated. Remarkable
shell structure emerges for certain combinations of quadrupole and
octupole deformations. Semiclassical periodic-orbit analysis indicates
that bifurcation of equatorial orbits plays an important role in the
formation of this new shell structure.
\end{abstract}
\journal{Physics Letters B}
\date{February 3, 1997}
\end{frontmatter}

Theoretical and experimental exploration of reflection-asymmetric
deformed shapes is one of the current topics of interest both in
nuclear-structure and micro-cluster
physics\cite{aberg,butler,brack,heiss}.

In theoretical calculations, various approaches like
Hartree-Fock-Bogoliubov methods, microscopic-macroscopic methods and
semi-classical methods have been exploited for this aim (see
Ref.~\cite{aberg} for a review).  Each method possesses merits and
demerits, so that it would be desirable to explore the subject with
various approaches.

A basic motive of the semi-classical periodic-orbit
approach\cite{gutw,balian,stru} is to understand the origin of
shell-structure formation that play decisive role in bringing about
symmetry-breaking in the average potential of finite quantum systems
like nuclei and micro-cluster.  If we could understand the origin and
obtain a global perspective, it would become possible to qualitatively
predict where we can expect a particular deformation to appear in the
multi-dimensional space spanned by various deformation-parameters and
the number of constituents of the system.

In the conventional wisdom, shell-structure would be weakened if
reflection-asymmetric deformation is added to the spheroidal shape.
This is because the system becomes non-integrable when the octupole
deformation is added, and because the degeneracy of the
periodic-orbits is then reduced.  Contrary to this expectation, a
significant shell structure was found in Refs.~\cite{arita,AM} to
emerge for certain combinations of quadrupole and octupole
deformations in the reflection-asymmetric deformed oscillator
model. It was pointed out that this shell-structure enhancement is
associated with bifurcation of periodic orbits.

It would be very important to investigate whether such a mechanism of
shell-structure formation is special to the harmonic-oscillator model
or possesses more general significance.

To explore the possibility that significant shell structure emerges in
the single-particle spectra for non-integrable Hamiltonian, we have
carried out an analysis of single-particle motions in the
reflection-asymmetric, axially-symmetric deformed cavity by
parameterizing the surface as
\begin{equation}
 R(\theta)=R_0\left(\frac 1{\sqrt{ (\frac {\cos\theta}a)^2 +
                                   (\frac {\sin\theta}b)^2 }} 
  + a_3 Y_{30}(\theta)
\right),
\end{equation}
where $a$ and $b$ are related with the familiar quadrupole deformation
parameter $\delta$ (equivalent to $\delta_{\rm osc}$ in
Ref.~\cite{bm}) by $a=((3+\delta)/(3-2\delta))^{2/3}$ and
$b=((3-2\delta)/(3+\delta))^{1/3}$.  It reduces to spheroid
(integrable cavity) in the limit that the octupole deformation
parameter $a_3$ vanishes.

We solve the Schr\"odinger equation under the Dirichlet boundary
condition and evaluate the shell energy by means of the Strutinsky
method.  To effectively obtain a large numbers of eigenvalues as a
function of deformation parameters, we have examined four numerical
recipes; the plane-wave decomposition (PWD)\cite{heller}, the
spherical-wave decomposition (SWD)\cite{pal}, the boundary integral
method (BIM)\cite{BR,BW,li} and the coordinate-transformation method
(DIAG)\cite{moszk,blocki}.
The DIAG is the most effective method for near-spherical shape, but
not good for strongly deformed shape.  In SWD, PWD and BIM, eigenvalue
problem is converted to a search of the zeros of real function, minima
of positive function and zeros of complex function, respectively, and
we found SWD is most convenient for the present purpose.  Thus we
mainly use this method, sometimes cross-checking the results by other
methods.

\begin{figure}
\epsfxsize=\figwidth
\centerline{\epsffile{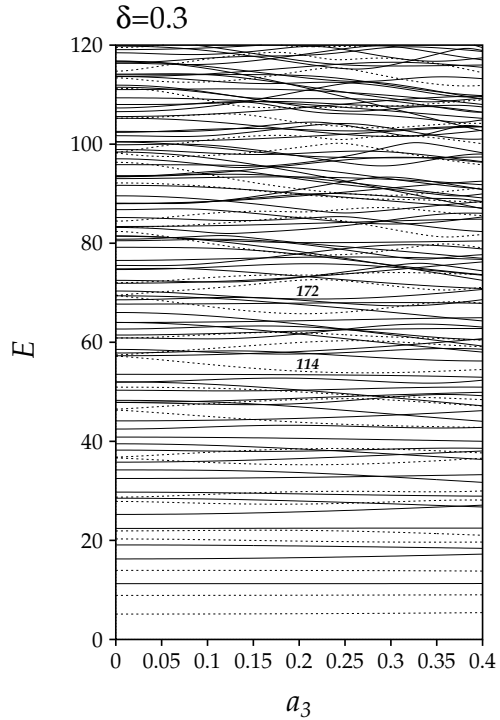}}
\caption{\label{fig1}
Single-particle-energy spectrum of the deformed cavity plotted as a
function of the octupole deformation parameter $a_3$.  The quadrupole
deformation parameter is fixed at $\delta=0.3$.  The energy is
measured in unit of $\hbar^2/MR_0^2$, $M$ being the mass.}
\end{figure}

Figure~\ref{fig1} shows a typical example of the single-particle
spectrum calculated as a function of the octupole-deformation
parameter $a_3$ fixing the quadrupole-deformation parameter at
$\delta=0.3$.  It is seen that a new shell structure emerges at about
$a_3=0.2$.  Deformed magic numbers associated with this shell
structure are 26,42,70,114,172,$\ldots$, taking the spin degeneracy
factor into account.  Note that these numbers appear at intermediate
places between the magic numbers 20,58,92,138,186,$\ldots$ of the
spherical cavity.  This indicates that, due to the reflection-symmetry
breaking of the cavity, strong $\varDelta l=3$ mixing takes place among
levels with large orbital angular momenta $l$ in spherical major
shells.

\begin{figure}
\epsfxsize=\figwidth
\centerline{\epsffile{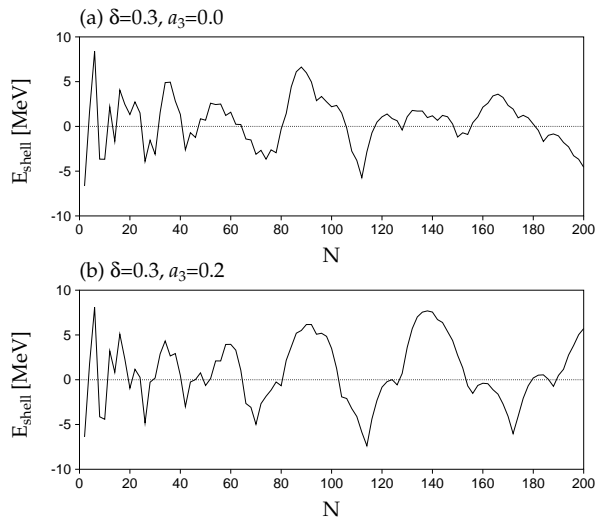}}
\caption{\label{fig2}
Shell structure energies of the deformed cavities with $\delta=0.3$
and $a_3=0.0$(a), $0.2$(b), evaluated with the conventional Strutinsky
method and plotted as functions of the particle number $N$.  The
energy is evaluated by putting $R_0=1.2(2N)^{1/3}$fm and
$Mc^2=938$MeV, for nuclei.}
\end{figure}

Figure~\ref{fig2} shows shell-structure energies evaluated with the
standard Strutinsky procedure and plotted as functions of the
particle number $N$. As expected from Fig.~\ref{fig1}, we confirm here
that minima develop in association with the formation of the new shell
structure at about $a_3=0.2$.

To understand the physical reason why such a remarkable shell
structure emerge for a certain combination of the octupole and
quadrupole deformations, and to identify the classical periodic orbits
responsible for this shell structure formation, we analyze Fourier
transform of the quantum spectrum.

The single-particle equations of motion for the cavity are
invariant with respect to the scaling transformation $(\vec{x},
\vec{p}, t) \to (\vec{x}, \alpha\vec{p}, \alpha^{-1}t)$.
The action integral $S_{\gamma}$ for the
periodic orbit $\gamma$ corresponds to the length $L_{\gamma}$ of it,
\begin{equation}
S_{\gamma}(E=p^2/2M)=\oint_{\gamma} \vec{p}\cdot d\vec{q}=pL_{\gamma},
\end{equation}
and the Gutzwiller trace formula is written as
\begin{eqnarray}
\rho(E)\simeq\bar{\rho}(E)+\sum_{\gamma}A_{\gamma}
k^{(d_{\gamma}-2)/2}\cos(kL_{\gamma}-\pi\mu_\gamma/2),
\nonumber\\
\end{eqnarray}
where $\bar{\rho}(E)$ denotes the contributions of orbits of `zero-
length', $d_{\gamma}$ the degeneracy and $\mu_{\gamma}$ the Maslov
phase of the periodic orbit $\gamma$.
This scaling property enables
us to make use of the Fourier transformation of the level density with
respect to the wave number $k$.  The Fourier transform $F(L)$ of the
level density $\rho(E)$ is written as
\begin{eqnarray}
F(L)&=&\int dk\, k^{-(d-2)/2}e^{-ikL}\rho(E=\hbar^2k^2/2M) \nonumber\\
    &\simeq&\bar{F}(L) + \sum_{\gamma}A'_{\gamma}\delta(L-L_{\gamma}).
\end{eqnarray}
which may be regarded as `length spectrum' exhibiting peaks at the
lengths of individual periodic orbits.  In numerical calculation, the
spectrum is cut off by Gaussian with cut-off wave number
$k_c=1/\varDelta L$ as
\begin{eqnarray}
F_{\varDelta L}(L)&\equiv&\int dk\, k^{-(d-2)/2}e^{-ikL}\,
e^{-\frac12(k/k_c)^2}
\nonumber\\
\noalign{\vspace{-5pt}}
&&\hspace{24mm}\times
\rho(E=\hbar^2k^2/2M) \nonumber\\
\noalign{\vspace{5pt}}
&&\hspace{-14mm}=
\frac{M}{\hbar^2}\sum_n k_n^{-d/2}e^{-ik_nL}\,
e^{-\frac12(k_n/k_c)^2} \label{eq:fourier-qm}\\
&&\hspace{-14mm}\simeq
\bar{F}_{\varDelta L}(L) + \sum_{\gamma}A''_{\gamma}
\exp\left[-\frac12\left(\frac{L-L_\gamma}{\varDelta L}\right)^2\right].
\label{eq:fourier-cl}
\end{eqnarray}
The amplitude $A_\gamma$ (or $A''_\gamma$) is proportinal to the
stability factor $1/\sqrt{|2-\Tr M_\gamma|}$ (in stationary-phase
approximation), where $M_\gamma$ is the monodromy matrix of orbit
$\gamma$, and expected to be enhanced in the vicinity of the
bifurcation point where $\Tr M_\gamma=2$ (see Ref.~\cite{AM}).

\begin{figure}
\epsfxsize=\figwidth
\centerline{\epsffile{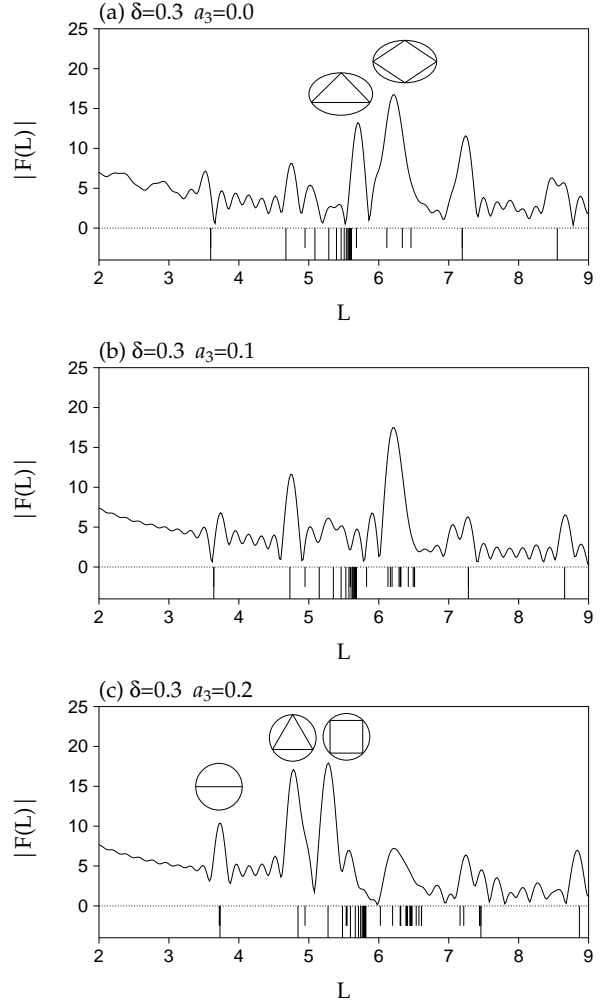}}
\caption{\label{fig3}
Fourier transforms of the quantum level densities for the deformed
cavities with $\delta=0.3$ and $a_3=0.0$(a), $0.1$(b), $0.2$(c).  The
degeneracy index $d=1$ (valid for generic periodic orbits) and
Gaussian cut-off wave number $k_c=\protect\sqrt{300}$ are used in
(\protect\ref{eq:fourier-qm}).  In each panel, lengths of classical
periodic orbits in the axis-of-symmetry (equatorial) plane are
indicated by short (long) vertical lines.  The lengths are measured in
unit of the radius $R_0$.}
\end{figure}

Let us investigate how these peaks change when the shape parameters of
the cavity is varied.  Figure~\ref{fig3} shows, as an example, how the
pattern of the Fourier transform (\ref{eq:fourier-qm}) changes as a
function of $a_3$, fixing the quadrupole deformation parameter at
$\delta=0.3$.  The highest peaks at the spheroidal limit ($a_3=0$) are
associated with triangular and quadrilateral orbits in the
axis-of-symmetry plane, whose degeneracies are two.  It is clearly
seen that these peaks decline with increasing $a_3$.  This is because
the octupole deformation breaks the spheroidal symmetry and the
degeneracy reduces to one corresponding to the rotation about the
symmetry axis.  On the other hand, we can clearly see that new peaks
rise with increasing $a_3$. These new peaks are found to be associated
with the periodic orbits in the equatorial plane at the center of the
larger cluster of the pear-shaped cavity.

\begin{figure}
\epsfxsize=\figwidth
\centerline{\epsffile{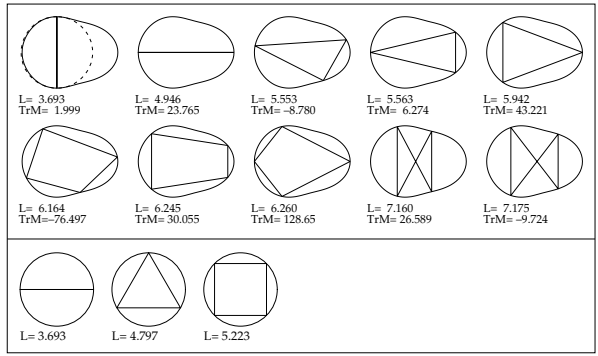}}
\caption{\label{fig4}
Periodic orbits in the deformed cavity with $\delta=0.3$ and
$a_3\simeq0.16$ (at bifurcation).  For each periodic orbit, the length
$L$ and the trace of the monodromy matrix, $\Tr M$, are indicated.
Those in the axis-of-symmetry plane are displayed in the upper panel
and those in the equatorial plane in the lower panel.  Only linear,
triangular and quadrilateral orbits are displayed.  In the
top-leftmost figure, a sphere tangent to the boundary at equatorial
plane is indicated by a broken line.}
\end{figure}

\begin{figure}
\epsfxsize=\figwidth
\centerline{\epsffile{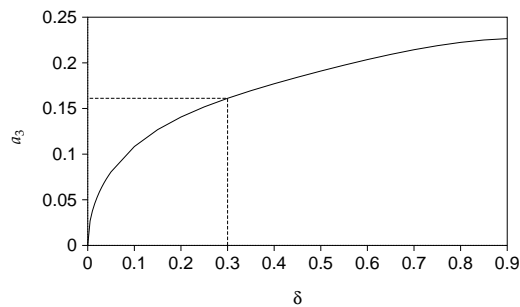}}
\caption{\label{fig5}
Bifurcation line of the equatorial periodic orbits in the
quadrupole-octupole deformation parameter space.  For $\delta=0.3$,
bifurcation occurs at $a_3\simeq 0.16$.}
\end{figure}

The key to understand the reason why such equatorial orbits start to
play increasingly important roles at finite octupole deformation may
lie in the following point: At certain combinations of $\delta$ and
$a_3$, the curvature radius of the cluster matches with its radius, as
illustrated in Fig.~\ref{fig4}.  This means that the phase space
structure around these periodic orbits {\em locally} becomes that of
`spherical' one.  Then bifurcation of these equatorial periodic orbits
occurs and new three-dimensional orbits are born.  Figure~\ref{fig5}
shows the bifurcation lines of this kind in the quadrupole-octupole
deformation parameter space.  The bifurcation occurs at $a_3\simeq
0.16$ for the case $\delta=0.3$.  In general, remarkable shell
structure may appear along the bifurcation line and stabilizes the
reflection-asymmetric deformed shapes.

In conclusion, we have investigated shell structure of the
single-particle spectrum in reflection-asymmetric deformed cavity.  It
is found that remarkable shell structure emerges for certain
combinations of quadrupole and octupole deformations.  Semiclassical
periodic-orbit analysis has been carried out, and it is found that
bifurcation of periodic orbits in the equatorial plane plays an
important role in the formation of this new shell structure.

A more detailed analysis of the equatorial-orbit bifurcation will be
reported elsewhere.

\section*{Acknowledgments}

We thank Matthias Brack and Alexander Magner for friendly
correspondences on the subject discussed in this letter.  Recently,
they have also found that the equatorial orbits play an important role
in asymmetric fission process (private communication).
We also thank Zhang Xizhen, Rashid Nazmitdinov and Masayuki Matsuo for
stimulating conversations.


\begin{thebibliography}{99}
\bibitem{aberg}
S. {\AA}berg, H. Flocard and W. Nazarewicz,
Ann. Rev. Nucl. Part. Sci. 40 (1990) 439.
\bibitem{butler}
P.A. Butler and W. Nazarewicz, 
Rev. Mod. Phys. 68 (1996) 349.
\bibitem{brack}
M. Brack, S. Creagh, P. Meier, S. Reimann and M. Seidl,
to be published in Proc. of the NATO ASI
{\it Large Clusters of Atoms and Molecules},
Erice 1995, edited by T.P. Martin.
\bibitem{heiss}
W.D. Heiss, R.G. Nazmitdinov and S. Radu,
Phys. Rev. B 51 (1995-I) 1874.
\bibitem{gutw}
M.C. Gutzwiller,
J. Math. Phys. 12 (1971) 343.
\bibitem{balian}
R. Balian and C. Bloch,
Ann. Phys. 69 (1972) 76.
\bibitem{stru}
V.M. Strutinsky, A.G. Magner, S.R. Ofengenden and T. D{\o}ssing,
Z. Phys. A 283 (1977) 269.
\bibitem{arita}
K. Arita,
Phys. Lett. B 335 (1994) 279.
\bibitem{AM}
K. Arita and K. Matsuyanagi,
Nucl. Phys. A 592 (1995) 9.
\bibitem{bm}
A. Bohr and B.R. Mottelson,
{\it Nuclear Structure} (Benjamin, 1975) Vol. 2.
\bibitem{heller}
E.J. Heller,
in {\it Chaos and quantum systems},
{\it Proc. NATO ASI Les Houches Summer School}, 1991,
edited by M-J. Giannoni, A. Voros and J Zinn-Justin
(North-Holland) p.547.
\bibitem{pal}
T. Mukhopadhyay and S. Pal,
Nucl. Phys. A 592 (1995) 291.
\bibitem{BR}
M.V. Berry and M. Robnik,
J. Phys. A: Math. Gen. 19 (1986) 649.
\bibitem{BW}
M.V. Berry and M. Wilkinson,
Proc. R. Soc. Lond. A 392 (1984) 15.
\bibitem{li}
B. Li and M. Robnik,
J. Phys. A: Math. Gen. 27 (1994) 5509.
\bibitem{moszk}
S.A. Moszkowski,
Phys. Rev. 99 (1955) 803.
\bibitem{blocki}
J. Blocki, J. Skalski and W.J. Swiatecki,
Nucl. Phys. A 594 (1995) 137.
\end{thebibliography}
\end{document}